\documentclass[twocolumn,10pt]{article}

\usepackage[english]{babel}
\usepackage{amsmath}
\usepackage{amsfonts}
\usepackage{amsthm}
\usepackage{amssymb}
\usepackage{rotating}
\usepackage{graphicx}
\usepackage{subcaption}
\usepackage{color}
\usepackage{hyperref}
\usepackage{booktabs}
\usepackage{authblk}
\hypersetup
 {  colorlinks=true,
    linkcolor=red,
    filecolor=red,      
    urlcolor=red,
    citecolor=red
    }

\newcommand{\bA}{\mathbf{A}}
\begin{document}

\title{Leveraging local network communities to predict academic performance}

\author[1,2]{David Burstein}
\author[3]{Franklin Kenter}
\author[4]{Feng Shi\thanks{bill10@email.unc.edu}}
\affil[1]{Swarthmore College}
\affil[2]{Fifth Third Bank}
\affil[3]{United States Naval Academy}
\affil[4]{University of North Carolina at Chapel Hill}

\date{}

\maketitle

\abstract
For more than 20 years, social network analysis of student collaboration networks has focused on a student's centrality to predict academic performance.  And even though a growing amount of sociological literature has supported that academic success is contagious, identifying central students in the network alone does not capture how peer interactions facilitate the spread of academic success throughout the network. Consequently, we propose novel predictors that treat academic success as a contagion by identifying a student's learning community, consisting of the peers that are most likely to influence a student's performance in a course. We evaluate the importance of these learning communities by predicting academic outcomes in an introductory college statistics course with $103$ students.  In particular, we observe that by including these learning community predictors, the resulting model is $68$ times more likely to be the correct model than the current state-of-the-art centrality network models in the literature. 





\section{Introduction} 
Education research has long acknowledged that peer interactions play a critical role in students' academic success \cite{dennis2005role,johnson2007state,johnson2014cooperative,mccabe2016friends}.  Consequently, to accurately identify academically at-risk students and implement innovative pedagogic strategies in the classroom, we need to incorporate our understanding of the network of student interactions, or student collaboration network, to inform such decisions.  The transformative potential for network analysis is not exclusive to education, but has influenced many other disciplines as well including neuroscience and epidemiology. In particular, epidemiologists have employed network analysis to design vaccination strategies for preventing the spread of biological contagions \cite{christakis2010social,cohen2003efficient}. Further, network science has shown that intangible characteristics such as happiness and sharing information on social media permeate like tangible pathogens \cite{fowler2008dynamic,hodas2014simple}. We anticipate that the successes of network science in analyzing biological contagions can be transferred to educational applications as well, since academic success is also contagious \cite{blansky2013spread}.  

Prior work using social network analysis has primarily focused on leveraging the student's position, or centrality, in the network to predict academic outcomes. \cite{baldwin1997social,bruun2013talking,grunspan2014understanding}.  While network centrality has been shown to be a statistically robust predictor of academic performance, such analyses do not take into account the heterogeneity of learning connections in the network. Without integrating network measures that account for the learning process on the student collaboration network, we cannot expect to develop accurate models to predict academic success.  Consequently, by identifying attributes important to the learning process, and measuring these attributes in context to a student's community, consisting of the peers that are most likely to influence a student's performance, we more faithfully model academic success as a social contagion process and achieve significant improvements over existing models. 

\section{Data}
We collected student data from an introductory statistics course with 103 students. Each student listed their collaborators on every graded assignment throughout the semester. The course contains units in probability, hypothesis testing and confidence intervals, where students apply their understanding of the material to analyze real-world data sets using the {\it R} statistical software language.

Throughout the course, there were six graded collaborative assessments, consisting of five homework assignments and one in-class assignment. When submitting homework assignments, students had the opportunity to fill out a short questionnaire identifying their collaborators for the assignment, their interest in the material on a Likert scale, and their confidence in the same manner. While some students did not complete all of the questionnaires, all 103 students agreed to participate in the study. 

Additional background information pertaining to their prior quantitative (i.e., mathematics and statistics) coursework and college entrance exams scores was collected at the beginning of the semester, since the results of these exams have correlated with students' first-year performance in college \cite{satpredict}. For the {\it SAT} entrance exams, only the general math score is used. Since some students took the {\it ACT} instead of the {\it SAT}, we converted {\it ACT} scores by identifying the corresponding {\it SAT} score in the same percentile. \cite{dorans1999correspondences}.

\subsection{Network Construction}

\begin{figure*}
\centering 
\includegraphics[width=\textwidth]{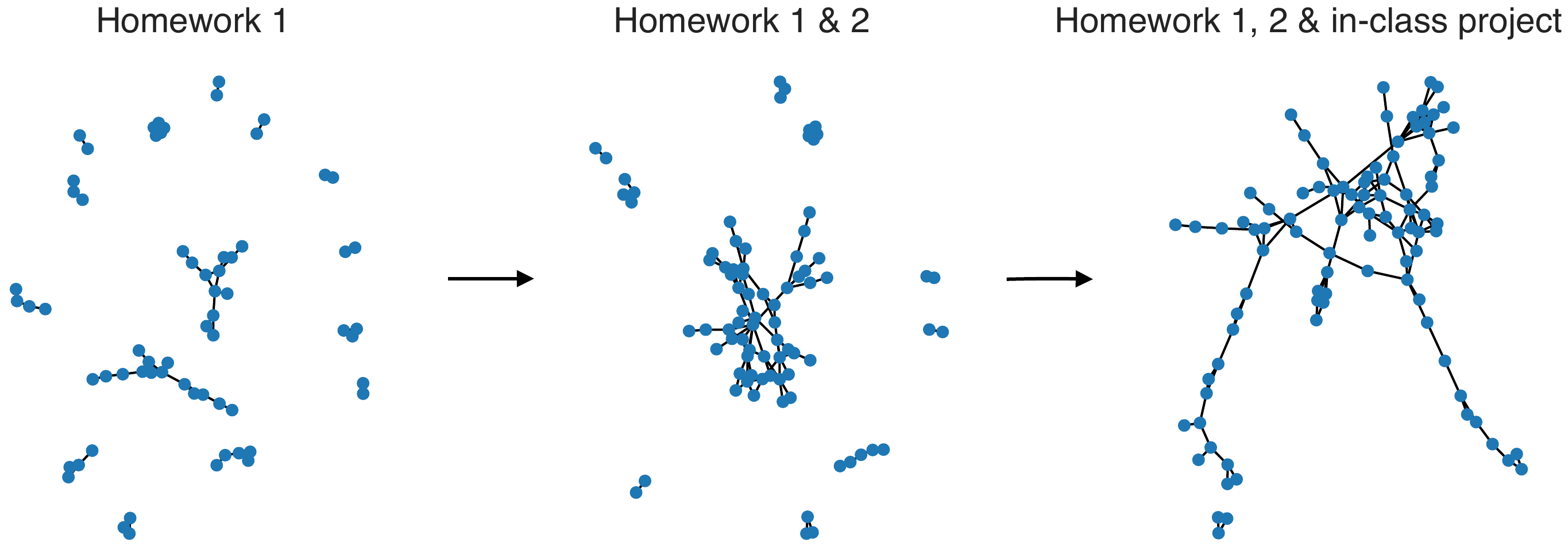}
\caption{Visualization of the student collaboration network constructed from homework 1 (left), homework 1 and 2 (middle), and homework 1, 2 and the in-class project (right). Nodes represent the students. An edge exists between two nodes if any of the two ever reported collaboration with the other in any of the assignments used to construct the network. }
\label{fig.snapshots}
\end{figure*}

The collaboration network of students consists of one node for each student with an edge from student $x$ to student $y$ if either $x$ or $y$ reported that they collaborated on a given assignment. For the purposes of the network, we consider the collaborative relationships to be symmetric and only consider undirected networks. Since there were 6 assignments, we built 6 collaboration networks cumulatively with each network built on top of the prior one. 
Specifically, for assignment $i$, we consider the network generated by all of the collaboration data up to and including assignment $i$.
For emphasis, we are not concerned with how often students work with each other, but rather, only if there is a reported relationship and when that relationship begins. This process results in six consecutive and different unweighted networks, each with progressively more edges than its predecessor. Snapshots of this network can be found in Figure \ref{fig.snapshots}.

\section{Centrality Measures}

\begin{figure}[!htb]
\centering
\includegraphics[width=0.2\textwidth]{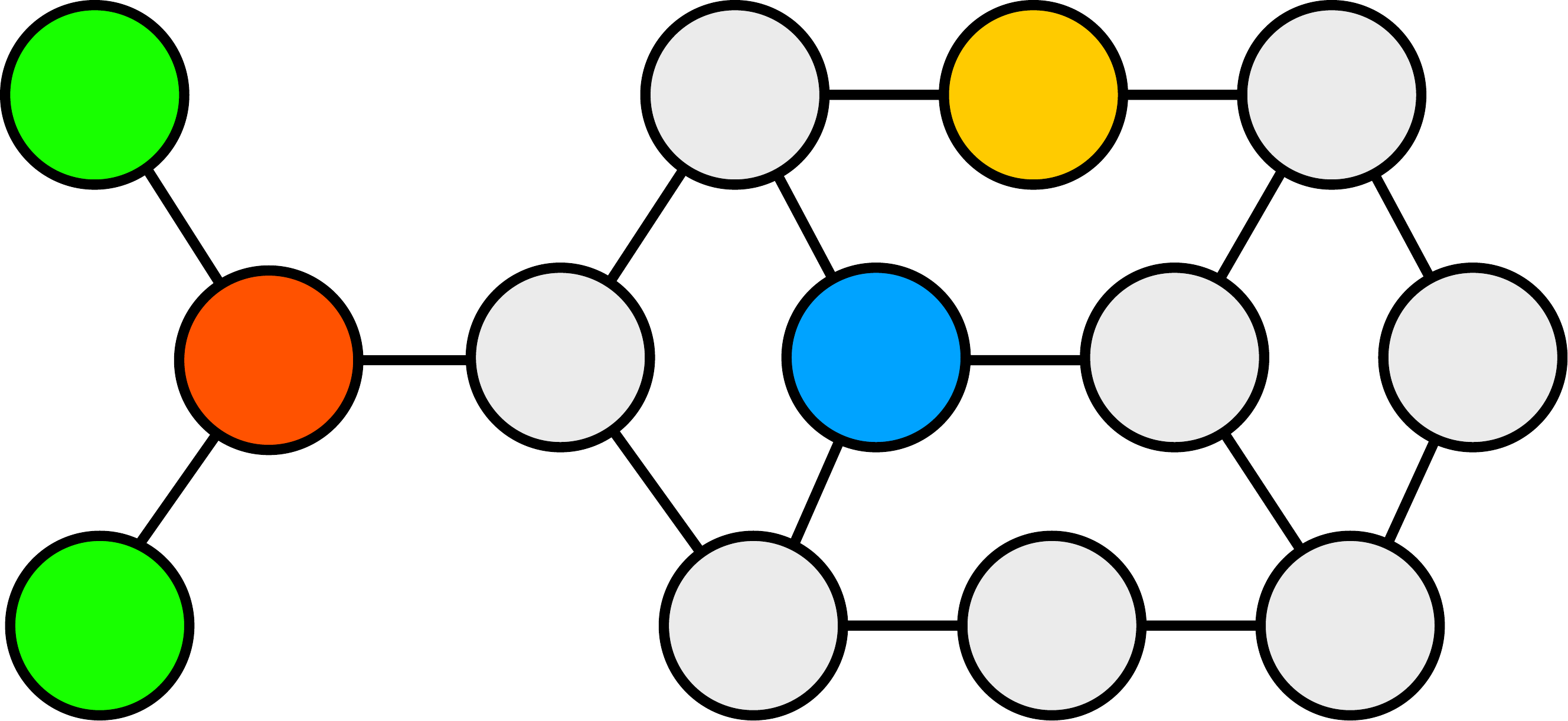}
\caption{The notion of centrality is mathematically \newline ambiguous.   The red and blue nodes have the same \newline degree-centrality but different target entropy centrality.}
\label{fig:centrality}
\end{figure}

While much prior literature has emphasized the importance of integrating tools from network science to predict academic outcomes, such analysis has been primarily restricted to measuring the centrality of nodes in the network \cite{ gavsevic2013choose, putnik2016analysing, thomas2000ties}.  In the context of student collaboration networks, students with high centrality are well-connected with their peers.  Since there are many mathematical definitions of centrality, we briefly review some prominent measures of centrality and later use these measures as a benchmark to evaluate the significance of our newly proposed community-based network measures for modeling academic performance. In most cases, technical definitions can be found in the Appendix.

\subsection{Degree and Log-Degree}

The simplest centrality measure is the {\it degree} of a node which identifies the number of connections emanating from that node. Observe that in Figure \ref{fig:centrality}, the blue and red nodes both have a degree of $3$, while the yellow node has a degree of $2$.  In the student collaboration network, the degree of a student indicates the number of other students they collaborated with whether reported by themselves and/or another student. Consequently, we anticipate that students who have many collaborators are more likely to succeed in the course. However, since studying with more student would likely have diminishing returns (i.e., having twice as many collaborators is not twice as good), we instead take the logarithm of the degree, or {\it log-degree}, to measure centrality. Rigorously, we denote the log-degree centrality of student $i$ as $\log(d_i)$, where $d_i$ denote the degree of a student $i$.

\subsection{Target entropy}
While degree centrality provides a straightforward method for identifying central nodes in the network, degree centrality implicitly treats all connections equally, even though certain connections are more valuable than others.  In the context of student collaboration networks, each  connection a student possesses provides additional access for that student to learn from their peers; nevertheless, certain connections are less likely to promote learning than others. In Figure \ref{fig:centrality}, the green nodes next to the red node are not well-connected to the rest of the network and do not provide additional access to red node's peers in the network.  As a result, the red node effectively only has one connection that grants access to the network. In contrast, all of the connections of the blue node increase its accessibility in the network.  Consequently, even though the red and blue nodes have the same degree centrality, the blue node has a larger {\em target entropy}.  Since a larger target entropy indicates that a student has many valuable connections that enhance its accessibility in the network, and hence more diverse resources to learn from their peers, we anticipate that target entropy should be correlated with academic success \cite{bruun2013talking}. We readily note that many other centrality measures have also been studied in context to student collaboration networks and provide a more technical discussion of these measures along with target entropy in the Appendix.  

\section{Community measures}

\begin{figure}[!htb]
\centering
\includegraphics[width=0.3\textwidth]{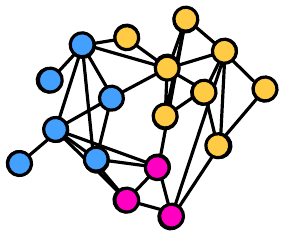}
\caption{An illustration of communities in a hypothesized network. There  are  three disjoint communities highlighted in  pink,  blue  and  yellow, with denser connections within communities than between communities.}
\label{fig:community1}
\end{figure}

As mentioned previously, centrality does not inherently capture academic success as a social contagion process.  Since we anticipate that academic success spreads throughout the network, we hypothesize that we can more accurately predict academic outcomes by identifying groups of students that are more likely to influence one another.  We call these groups of nodes {\em communities}.  Mathematically, this suggests that nodes within the same community are more likely to be connected with one another.  Consider the network in Figure \ref{fig:community1}, where there are three communities highlighted in pink, blue and yellow.  Note that the yellow nodes tend to be connected with one another.  While the mathematical description of a community may appear straightforward, there are in fact many definitions for a  community and many implementations for finding them \cite{lancichinetti2009community}. We describe two of these methods, {\em Personalized Pagerank} and {\em Infomap} below. 

\subsection{Personalized Pagerank}
Personalized Pagerank identifies a community for each node by performing a random walk on the network \cite{andersen2006local}.  Without loss of generality, below we describe the procedure for constructing a personalized community for a focal node $v$. Suppose we are currently located at any given node in the network. With probability $1-\alpha$ we move to a node picked uniformly at random from all the nodes connected to the current one, and with probability $\alpha$ we teleport to $v$. Through simulating many time steps, we attain probabilities that we visit any node in the network when $v$ is the focal node.  Accordingly, we use those probabilities as the weights of other nodes in $v$'s community. In other words, nodes with larger weights are more likely to be in $v$'s community.    

\begin{figure*}
\centering
\includegraphics[width=0.4\textwidth]{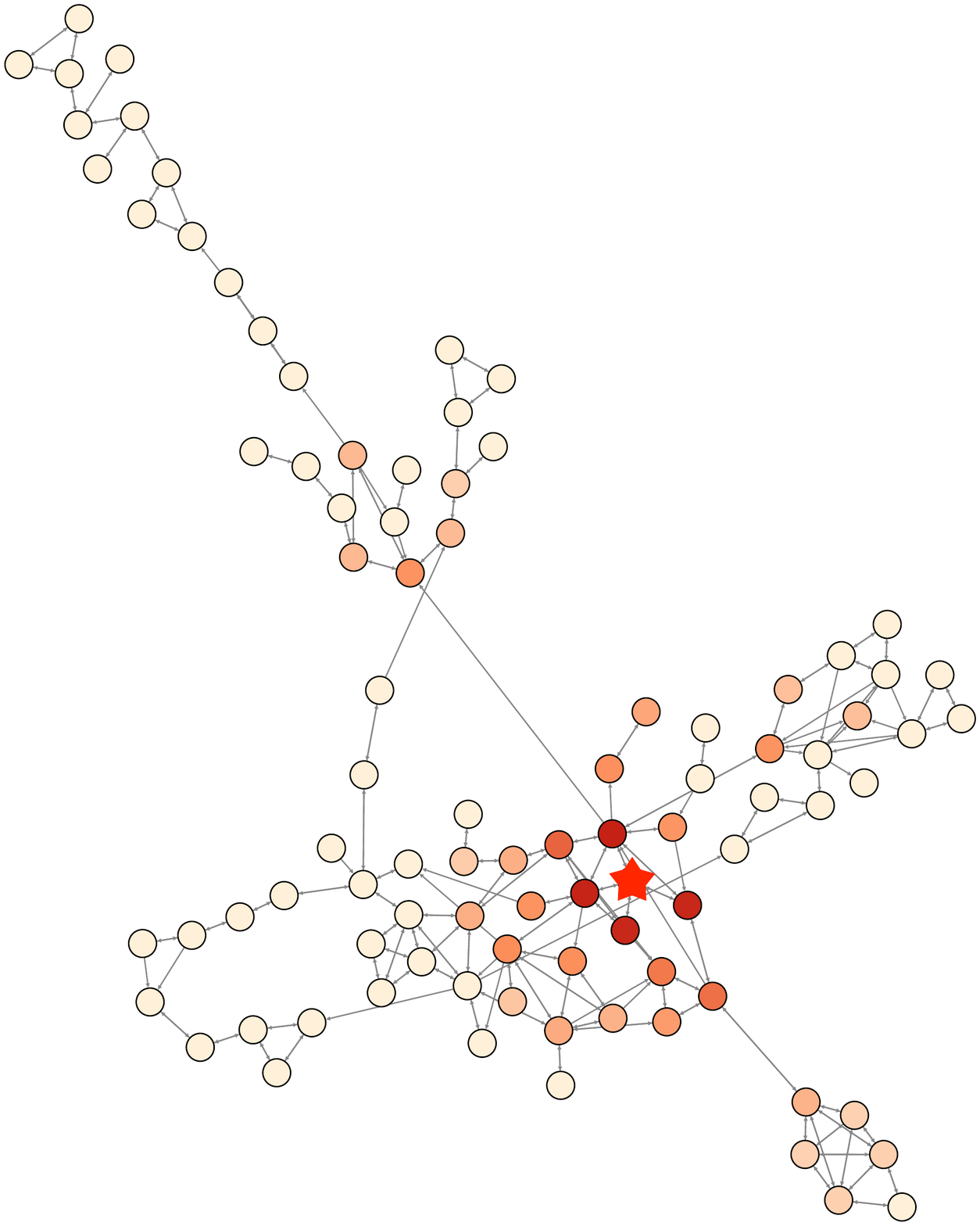} 
\includegraphics[width=0.4\textwidth]{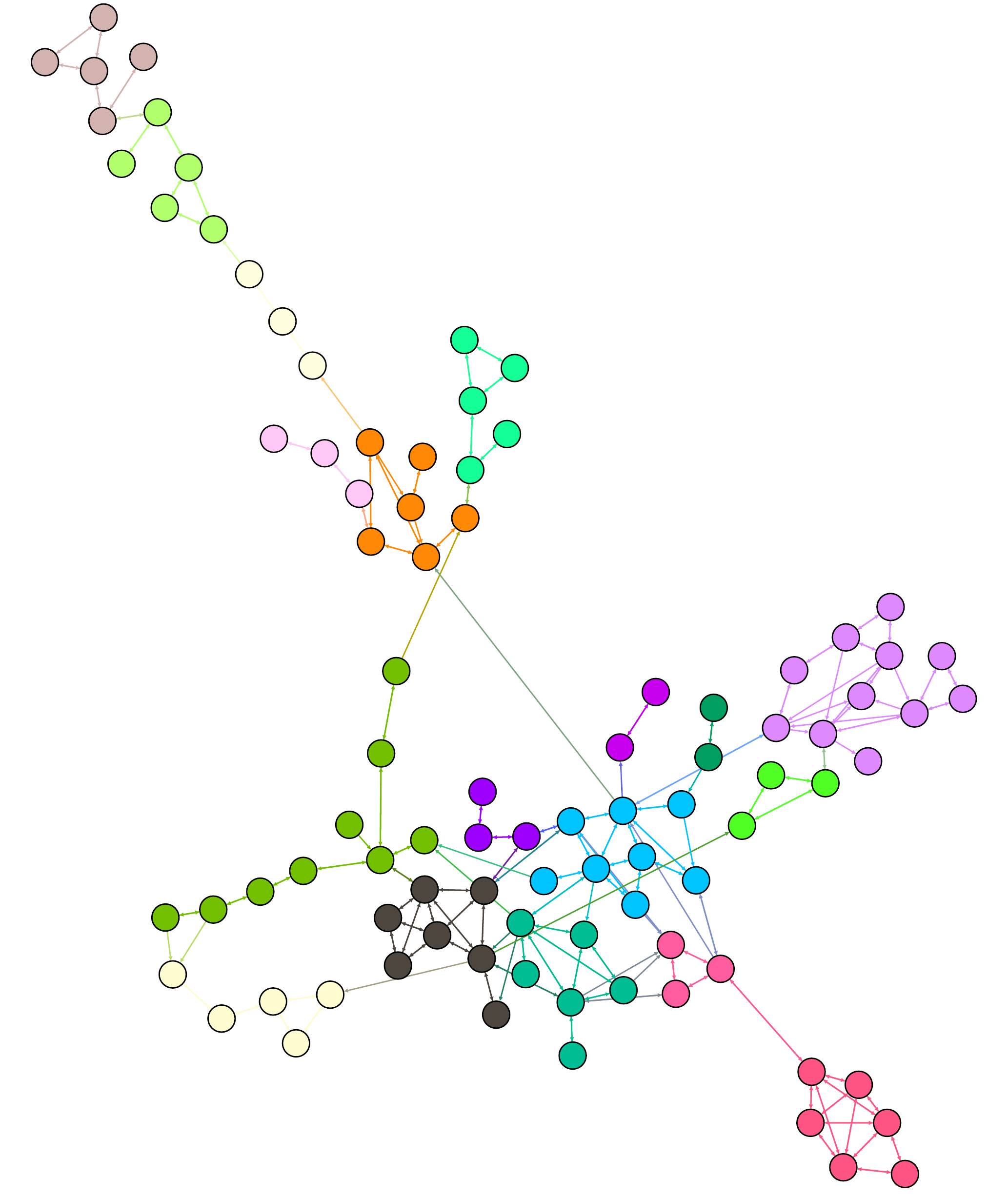}
\caption{Community structure in the student collaboration network constructed from all assignments. Left: The pagerank score of everyone in the star node's personalized community, with darker color corresponding to higher score. Right: Communities identified by the Infomap method, with colors denoting different communities. }
\label{fig:community}
\end{figure*}

In the left network displayed in Figure \ref{fig:community}, we computed the Personalized Pagerank for the star node, where nodes with darker shades of red have larger Pagerank scores and hence are more likely to be in the community of the star node. Recalling our informal definition of communities, observe that the darker nodes are more likely to share a connection with each other.  In fact, by including a rule that we teleport back to node $v$, we ensure that nodes with large weights are typically close to $v$.

At this juncture, we would like to emphasize that the output of this method produces a {\em personalized} community.  This means that nodes $u$ and $v$ can have large weights in each others' personalized community, but the personalized community for node $u$ and the personalized community for node $v$ can look different.  Furthermore, note that in contrast to Figure \ref{fig:community1}, where we partition nodes into fixed, disjoint clusters, Personalized Pagerank is an example of a soft partition.  We do not assign a threshold to determine whether a particular node is part of a community; the output of Personalized Pagerank only provides weights to illustrate if a node is relatively suited to be part of another node's community. 

After attaining the list of weights for each node's personalized community, we can then construct community-weighted measures.  For example, instead of looking at an individual student's self-rated confidence in the material, we can use these weights to take a weighted average of the confidence of that student's personalized community, consisting of the peers that are most likely to influence that student's performance. When computing this weighted-average, we exclude that student from the computation. Naturally, we can  apply the same technique to create other predictors as well, including  community-averaged interest scores and   community-averaged SAT scores.  By designing predictors that take into account the attributes of a student's peers that are most likely to influence their academic performance, we are building a predictive model that is able to exploit the contagious nature of academic success.  

\subsection{Infomap Community Detection}

We emphasize that there are {\em many} methods for detecting communities. However,  the effect of student communities, if exists, should not be solely determined by the method used to find the communities. Hence, we will utilize another prominent community detection method, Infomap \cite{bohlin2014community}, to ensure the robustness and the consistency of the community-based effects. 

Infomap finds communities such that typical paths can be expressed in an efficient manner. Similar to how highly-trafficked roads have short names (i.e., short highway numbers), Infomap aims to label nodes so those involved in many paths have shorter names. As with roads names, labels for Infomap can be repeated in different communities. However, the communities for Infomap are not predefined. Rather, Infomap chooses its communities based on how effectively it can choose the labels within each of them. 

\begin{figure}
\centering
\includegraphics[width=0.4\textwidth]{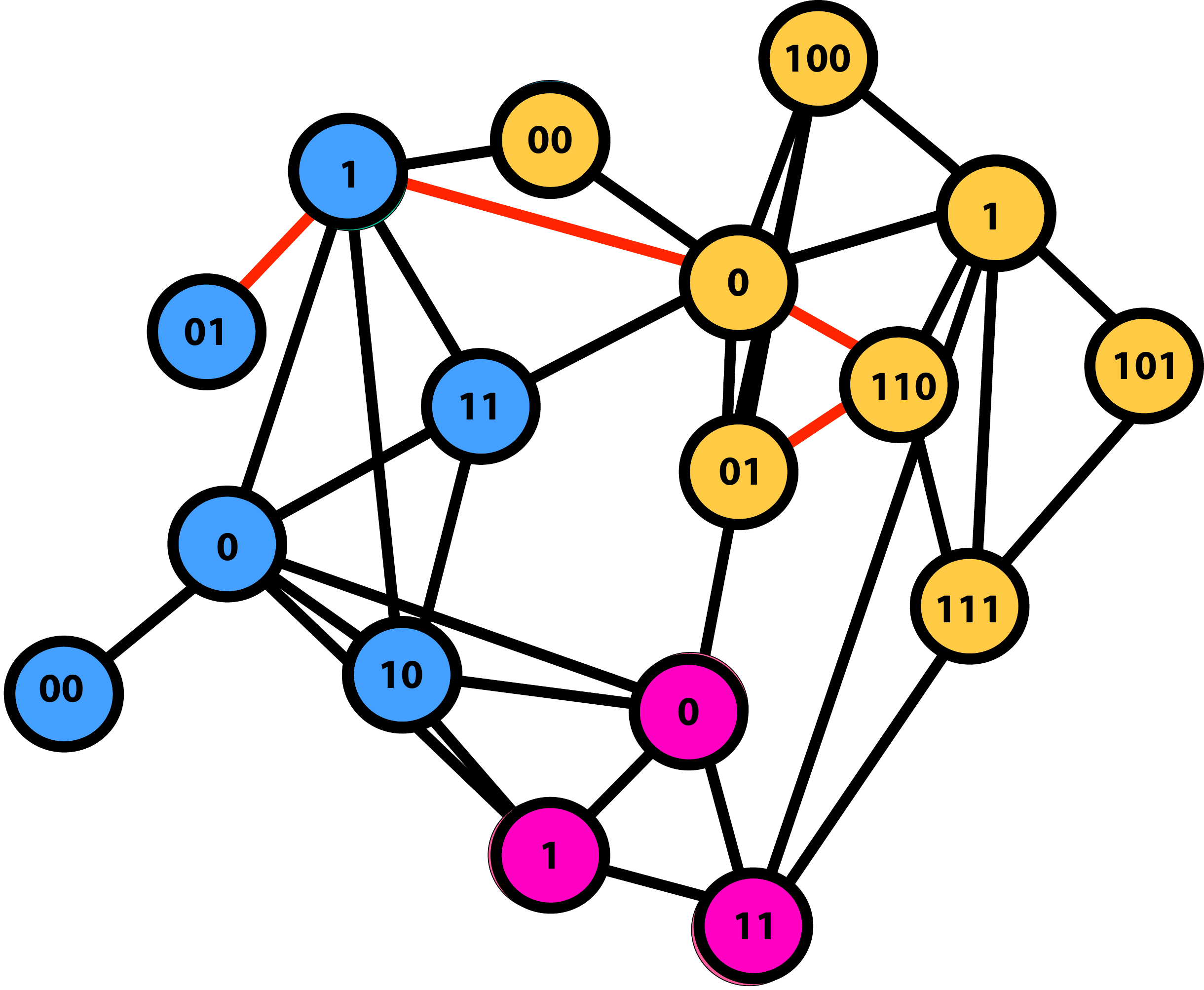}
\caption{An illustration of the Infomap method. The network has three communities denoted by colors. Each node is coded using a  binary  numeral system, and nodes that are visited frequently by random walks are given shorter names. For example, the blue node 0 is more central than the blue node 00 and hence has a shorter name. The example red path can be expressed as ${\color{blue}0},01,1,{\color{blue}11},{\color{yellow}1},0,110,01,{\color{yellow}00}$ with code words highlighted in blue and yellow to indicate that the path is entering or leaving a community.}
\label{fig:infomaplocal}
\end{figure}

For an example, in Figure \ref{fig:infomaplocal}, we name each node using a binary numeral system.  For example, the path highlighted in red can be written as $01,1,0,110,01.$ 
To express paths efficiently, nodes that are visited frequently are given shorter names.  From Figure \ref{fig:infomaplocal} the blue node $0$ is more central than the blue node $00$ and hence has a shorter name.  We also allow reusing the same name for nodes in different communities, much like street names are reused in different cities, by including code words indicating that we have in fact changed communities. 
Consequently, to express the path above, we add code words highlighted in blue and yellow to indicate that we are entering or leaving a community.
$${\color{blue}0},01,1,{\color{blue}11},{\color{yellow}1},0,110,01,{\color{yellow}00}.$$
While it might appear advantageous to have many communities, so that we can reuse short names for nodes many times, we also incur a penalty when we transition between communities. Hence, when using Infomap, we must balance the the trade-off of using small communities with many transitions and large communities with longer node names.  

For emphasis, Infomap only outputs the optimal community structure that corresponds to the optimal encoding of paths, not the encoding of the paths themselves.  

In connection to student collaboration networks, if two students are in the same Infomap community, that means there are many short paths between them, and hence these students are more likely to influence each other.  Consequently, we anticipate that the communities output by Infomap will provide predictive insight for modeling academic performance in the classroom.  

Now we can construct community-averaged measures for confidence, SAT scores, interest, etc., by applying the same procedure described in the Personalized PageRank section.  Finally, note that in contrast to  Personalized PageRank, which provides weights indicating the plausibility of a node belonging to a certain community, the Infomap algorithm treats community membership as a categorical variable. For this reason, we anticipate that Infomap might not perform as well as the Personalized PageRank, as it defines the community membership in a more restricted fashion; nevertheless, Infomap contributes a unique perspective in identifying non-trivial points of influence for student academic achievement.

\section{Results}
\begin{table}
\centering
\begin{tabular}{ |l|r|}
\hline
Predictor & Section\\
\hline
Confidence, Interest, and SAT & 2\\
Log-Degree & 3.1\\
Target entropy & 3.2\\
Hide & Appendix C\\
PageRank Community Confidence & 4.1\\
PageRank Community Interest & 4.1\\
PageRank Community SAT & 4.1\\
Infomap Community Confidence & 4.2\\
Infomap Community Interest & 4.2\\
Infomap Community SAT & 4.2\\
\hline
\end{tabular}
\caption{Reference table for definitions of the predictors used in the regression model.}
\label{tab:predictors}
\end{table}

We build a series of linear regression models to predict students' final exam scores with the community measures and other features of students as predictors. Specifically, we have three groups of predictors: (1) individual information including SAT scores and self-reported confidence and interest in the course; (2) network centrality measures including degree, target entropy and hide; (3) community-based measures including community-averaged SAT, confidence and interest. See Table \ref{tab:predictors} for a complete list of the predictors.

Accordingly, for the network at each time point, we estimate 3 families of models. The first family of models use only the first group of predictors (i.e., individual information), and serve as the baseline of prediction. The second family of models add centrality measures to the first and use both the first and the second group of predictors. They represent the state-of-the-art in predicting academic performance from student networks  \cite{bruun2013talking, grunspan2014understanding,thomas2000ties}. Finally, the third family of models incorporate the community-averaged measures into the second and use all the predictors. A brief summary for which predictors are included in each of the three families of models is shown in Table \ref{tab:whichpredictor}.

\begin{table*}[]
\centering
\begin{tabular}{lcccc}
\hline
                    & \multicolumn{1}{l}{}                & \multicolumn{1}{l}{}                  & \multicolumn{2}{c}{Community Models}                       \\
Predictors          & \multicolumn{1}{l}{Baseline Models} & \multicolumn{1}{l}{Centrality Models} & \multicolumn{1}{l}{Pagerank} & \multicolumn{1}{l}{Infomap} \\ \hline
Confidence          & x                                   & x                                     & x                            & x                           \\
Interest            & x                                   & x                                     & x                            & x                           \\
SAT                 & x                                   & x                                     & x                            & x                           \\
Log-Degree          &                                     & x                                     & x                            & x                           \\
Target Entropy      &                                     & x                                     & x                            & x                           \\
Hide                &                                     & x                                     & x                            & x                           \\
Pagerank Confidence &                                     &                                       & x                            &                             \\
Pagerank Interest   &                                     &                                       & x                            &                             \\
Pagerank SAT        &                                     &                                       & x                            &                             \\
Infomap Confidence  &                                     &                                       &                              & x                           \\
Infomap Interest    &                                     &                                       &                              & x                           \\
Infomap SAT         &                                     &                                       &                              & x                           \\ \hline
\end{tabular}
\caption{Quick look-up table for which predictors are included in each the three families of models.}
\label{tab:whichpredictor}
\end{table*}

We then compare the different models to assess the empirical significance of the community measures.  In order to evaluate community (or network) predictors, we only create predictions for students that have connections in the network.  Since some students did not report connections until later in the course, models corresponding to earlier time shots would not generate predictions for those students.  Furthermore, we cannot compare the $R^2$ of those models directly since adding more predictors will only increase the prediction accuracy. Instead, we use Akaike information criteria (AIC) as the measure of model performance which penalizes models with more predictors \cite{burnham2004multimodel}. Hence, for each family of models we perform a model selection based on their AIC's and report the one with the optimal AIC. If the community measures do not sufficiently improve the predictive performance, then they will not appear in the optimal model from the third family of models. 

We report the predictive performance of each of the three families of models in Table \ref{table:models} consisting of the $R^{2}$ statistic and the evidence ratio between the overall optimal model and each model listed in the table. Specifically, the evidence ratio compares the likelihood of the overall optimal model with the likelihood of the optimal model from each family listed, based on their AIC scores. Therefore, smaller numbers of the evidence ratio indicate better performance, and a ratio of 1 means that the listed model is as good as the overall optimal model.

\begin{table*}
\centering
\begin{tabular}{@{}lccclccc@{}}
\toprule
& \multicolumn{3}{c}{$R^2$} &  & \multicolumn{3}{c}{Evidence Ratio} \\
& HW 1 & HW 2 & In-Class &  & HW 1     & HW 2     & In-Class     \\ \cmidrule(lr){2-4} \cmidrule(l){6-8} 
Baseline Model           & .117 & .161 & .225     &  & 1.00     & 8.56     & 560.39       \\
Centrality Model         & .126 & .187 & .276     &  & 1.95     & 6.39     & 67.56        \\
Infomap Community Model  & .138 & .261 & .345     &  & 1.26     & 1.00     & 4.87         \\
PageRank Community Model & .133 & .246 & .367     &  & 1.51     & 2.27     & 1.00         \\ 
\bottomrule
\end{tabular}
\caption{Predictive performance of each of the three families of models. The evidence ratio is the ratio between the likelihood of the overall optimal model and the likelihood of the optimal model from each family listed, based on their AIC scores. Therefore, smaller numbers of the evidence ratio indicate better performance, and a ratio of 1 means that the listed model is as good as the overall optimal model.}
\label{table:models}
\end{table*}

From the evidence ratios listed in the column for Homework 1 (HW 1), we note that while the network-based models appear competitive with the baseline (non-network) model, the network models do not yield a substantial improvement over the baseline model.  We attribute this result to a lack of network data, since the average number of connections per node is close to 1 and the network is too sparse. 

After attaining the homework 2 collaboration data, the average number of connections per node is close to 2 and the network is significantly more connected.  Consequently,  adding community predictors yields a $62\%$ increase in $R^{2}$ from the baseline model and a $40\%$ increase in $R^{2}$ from the centrality model.  The evidence ratios also begin to demonstrate the improvement from the community predictors; for example, from the AIC evidence ratios, the optimal Infomap Community  model is the overall optimal model, and it is $8.65$ times more likely than the baseline models given our data. 

Since we conjecture that academic success spreads like a biological contagion, we anticipate that we can improve our predictions by including network data that captures how students interact both inside and outside the classroom.  When we include the connections from the in-class project, we attain statistically significant improvements by using community predictors.  In particular, the optimal PageRank Community Model is $560$ times more likely than the Baseline (non-network) Model and $68$ times more likely than the optimal Centrality Model, given our data.  Furthermore, these community predictors result in a $63\%$ increase in $R^{2}$ from the baseline model and a $33\%$ increase in $R^{2}$ from the centrality model.  We specify the coefficients for this optimal Pagerank Community Model in Table \ref{table:bestmodel}.  In particular, we observe that even though the Interest variable does not appear in the optimal model, its community analogue, Pagerank Interest is present in the model. Furthermore, the $p$-values for both of the community predictors are substantially smaller than the centrality predictor, Target Entropy.  Consequently, our analysis emphasizes the importance of including network community predictors to build accurate models for predicting academic success.   

\begin{table*}
\centering
\begin{tabular}{rcccc} 
\toprule
& Coefficient & $95\%$ Confidence Interval & p-value \\ \cmidrule{2-4}
Confidence  & 4.46   & (2.18, 6.75) & .0002    \\
PageRank Confidence  & 20.06 & (8.75, 31.37) & .0006 \\
PageRank Interest & -13.76 & (-23.82, -3.70) & .0079   \\
SAT  & 0.05  & (0.02, 0.09) & .0021   \\
Target Entropy  & 5.42  & (0.94, 9.90) & .0184   \\ \bottomrule
\end{tabular}
\caption{The regression results for the optimal model using the Homework 1, 2 and in-class network data.}
\label{table:bestmodel}
\end{table*}

\section{Discussion and Conclusion}
We propose novel predictors for academic performance based on a student's learning communities, consisting of the peers that are most likely to influence a student's academic performance in a homework collaboration network. Using only the first two homework assignments and one in-class assignment, we find that the network community-based models are 560 times more likely than models that use non-network predictors alone, given our data. We also note that the network community-based model is 68 times more likely than the state-of-the-art network centrality-based models. As a result, the accuracy of the community-based model suggests the contagious nature of academic success on social networks, similar to other intangible properties such as happiness or obesity.

When building the aforementioned predictive models for academic performance, we deliberately avoided predictors that incorporate homework and exam grades, (which are often not available at the beginning of a course).  Since we suspect students' midterm grades influenced the self-rated confidence scores reported in Homework 3-5, we focused our analysis on the data collected from Homework 1-2 and the In-Class Assignment, which capture $85\%$ of the network connections reported throughout the study (Table \ref{table:network_growth}).  In particular, we observe that the midterm had a marginal impact on the formation of new homework collaborations.  On account of the rapid stabilization of the network, we can infer that the collaborations recorded from the in-class assignment, which took place after the midterm, were present before the midterm.   Consequently, the performance of the community network models highlight the value of using community-based predictors for detecting academically at-risk students early in the course, when grades are not available.
 
\begin{table*}
\centering
\begin{tabular}{lcccccc}
\toprule
	& Homework 1	&	Homework  2	&	Homework 3	&	Homework 4	& In-class project	& Homework 5 \\
\hline
Nodes &	103	&	103	&	103	&	103	&	103		& 103 \\
Edges	& 119	&	176	&	197	&	212	&	275	&	281 \\
\bottomrule
\end{tabular}
\caption{The growth of edges in the student collaboration networks.}
\label{table:network_growth}
\end{table*}

Since our work only establishes a correlation between academic performance and the community predictors, further analysis is necessary to measure the causal effect of learning communities on academic performance.   Furthermore, our work exclusively utilizes linear models to evaluate the utility of community predictors in modeling academic outcomes, even though there are many other models for predicting academic performance. Consequently, we look forward to the development of data-driven approaches that leverage both state-of-the-art machine learning techniques and network community analysis to inform pedagogic practices in the classroom.  With these community-based predictors, we anticipate that our work will provide instructors with new tools to quickly and accurately identify academically at-risk students.  

\bibliographystyle{plain} 
\bibliography{pednetworks}

\appendix

\section{PageRank Centrality}

PageRank was originally developed to rank webpages on the internet to determine which ones are more influential \cite{page1999pagerank}, and it is a special case of the eigenvector centrality. The eigenvector centrality of a network is given by the eigenvector $\mathbf{v}$ corresponding to the maximum eigenvalue of the adjacency matrix of the network. PageRank is an adjusted eigenvector with a tuning parameter (also known as jumping parameter) $\alpha$. For a strongly connected graph, it is the solution to the generalized eigenvector equation \[ \mathbf{x}_\alpha = \alpha \bA^T \mathbf{D}^{-1} \mathbf{x}_\alpha + (1-\alpha) \frac{\mathbf{1}}{n}\]  where $\mathbf{A}$ is the adjacency matrix, $\mathbf{D}$ is the diagonal degree matrix, and $\mathbf{1}$ is the all-one vector.

PageRank can also be defined by the following random process: Start at a random node, with probability $\alpha$ move to a random adjacent node; otherwise (with probability $1-\alpha$), jump to a random node. Then, repeat the process. The PageRank vector, $\mathbf{x}_\alpha$ is the vector  describing the limiting probability you will visit each vertex after a large number of steps. 

\section{Target Entropy}

For a given node $i$, imagine that each other node send a message to $i$, and the messages travel along shortest paths. If there are $N>1$ shortest paths between some node and $i$, split the message evenly so that each shortest path carries $1/N$ message. For each neighbor $j$ of $i$, let $m_{ij}$ be the number of messages $i$ received from $j$. The messages can be originated from $j$ or forwarded by $j$, and the number $m_{ij}$ is not necessarily a whole number since a message could be split between multiple shortest paths. Lastly, let $M_i$ be the total number of messages $i$ received, and $c_{ij}$ be $m_{ij}/M_{i}$, the fraction of messages from $j$. The target entropy of $i$ is defined as
$$
T_i = - \sum_{j\sim i} c_{ij} \log(c_{ij})
$$

It is easier to interpret the target entropy with a hypothetical example. Assume a student A who interacts with many other students who are highly connected in the network; and another student B who has equally many friends but most of her friends are only connected to her. Then A would have larger target entropy than B. What this means is that A has many sources of information while B, being the source of information among her community, does not have as many sources of information as A.

\section{Hide}
Hide describes how hard a node can be found within a network \cite{bruun2013talking}. A hard to find node is usually lacking interactions with other nodes, hence providing information on the constraints that the student might experience. In our context, A student with high hide score might not have the same access to ideas and information as a student with low hide.

To introduce the definition of hide, we first define the search information $S(i, j)$ from node $i$ to node $j$. Let $i \rightarrow j $ denote a shortest path from $i$ to $j$, and $k - k' \in i \rightarrow j$ be any consecutive two nodes on the path from $i$ to $j$. Then the probability of a random walk following the path  $i \rightarrow j $ can be calculated as 
$$
P(i \rightarrow j) = \prod_{k - k' \in i \rightarrow j } \frac{w_{k k'}} {s^{out}_{k}},
$$
where $w_{k k'}$ is the weight of edge $k-k'$ and $s^{out}_{k}$ is the out-strength of node $k$. The product is taken over all adjacent edges on the path.  

The search information from $i$ to $j$ is then defined as:
$$
S(i, j) = - \log_2(\sum_{i \rightarrow j } P(i \rightarrow j )),
$$
where the summation is taken over all shortest paths from $i$ to $j$. The search information quantifies how much information a random walk needs to go from $i$ to $j$. It $S(i, j)$ is large, then it is hard to find $j$ from $i$ since it requires a lot of information. 

Finally, the hide score $H_i$ of a node $i$ is defined as the average information needed to find $i$ from any other nodes:
$$
H_i = \frac{1}{N-1} \sum_{j \in V, j\neq i} S(j,i),
$$
where $N$ is the number of nodes in the networks, and $V$ is the set of nodes.

\section{Personalized Pagerank}
The {\it personalized Pagerank community} of a node, $v$ is given by solving a slightly different generalized eigenvector equation 
$$
\mathbf{x}^v_\alpha = \alpha \bA^T \mathbf{D}^{-1} \mathbf{x}^v_\alpha + (1-\alpha) \mathbf{1}_{v},
$$ where $\mathbf{A}$ is the adjacency matrix, $\mathbf{D}$ is the diagonal degree matrix, $\alpha$ is a predetermined jumping parameter, and $\mathbf{1}_v$ is the indicator vector for node $v$. The result is a probability distribution on all the nodes, ``centered'' around $v$.

It should be noted that in many cases the vector $\mathbf{x}^v_\alpha$ is heavily supported at node $v$. Further, if $v$ is isolated, the vector $\mathbf{x}^v_\alpha$ is precisely $\mathbf{1}_v$.

\section{Infomap}

Infomap uses a random walk on a network as a proxy for the information flow on the network,
and aims to find a partition of the nodes such that the description length of random walks on the network is minimized. As an analogy in geography, think of each community as a city and the nodes as street addresses. Then a random walk can be described by enumerating the names of all the city and street names visited by the random walk. This description has a certain length and this length can change if the city and street names change. Hence, the community detection problem is equivalent to solving an optimal coding problem. Mathematically,  the average description length of a single step of random walks on a network is given by
\begin{equation}
L(M) = q H(Q) + \sum_{i=1}^m p^i H(P^i),
\end{equation}
where $M$ is a given partition of the network, $q$ is the probability that the random walk switches communities on any given step, $H(Q)$ is the entropy of the community names, $H(P^i)$ is the entropy of the within-community movements, and $p^i$ is the fraction of within-module movements that occur in community $i$, plus the probability of exiting community $i$. In the end, the output of this method is the partition $M$ of nodes that minimizes this description length. 

\end{document}